\RequirePackage{fix-cm}
\documentclass[twocolumn]{svjour3}          
\smartqed  
\usepackage{graphicx}
\usepackage[margin=1in]{geometry}
\usepackage{fancyhdr}
\usepackage{natbib}
\usepackage{url}

\usepackage{lipsum}  
\usepackage{setspace}
\usepackage{amsmath}
\usepackage{amssymb}
\usepackage{bm}
\usepackage{booktabs}

\usepackage{listings}
\usepackage{xcolor}
\definecolor{codegreen}{rgb}{0,0.6,0}
\definecolor{codegray}{rgb}{0.5,0.5,0.5}
\definecolor{codepurple}{rgb}{0.58,0,0.82}
\definecolor{backcolour}{rgb}{0.95,0.95,0.92}
\lstdefinestyle{mystyle}{
    backgroundcolor=\color{backcolour},   
    commentstyle=\color{codegreen},
    keywordstyle=\color{magenta},
    numberstyle=\tiny\color{codegray},
    stringstyle=\color{codepurple},
    basicstyle=\ttfamily\footnotesize,
    breakatwhitespace=false,         
    breaklines=true,                 
    captionpos=t,                    
    keepspaces=true,                 
    numbers=left,                    
    numbersep=5pt,                  
    showspaces=false,                
    showstringspaces=false,
    showtabs=false,                  
    tabsize=2
}
 
\begin{document}

\title{fbst: An R package for the Full Bayesian Significance Test for testing a sharp null hypothesis against its alternative via the e-value
}

\titlerunning{fbst: An R package for the Full Bayesian Significance Test}        

\author{Riko Kelter
}


\institute{Department of Mathematics\at
              University of Siegen, Germany \\
              \email{riko.kelter@uni-siegen.de}             \\
             \emph{Present address:} Walter-Flex-Street 3, 57072 Siegen, Germany\\\
             \emph{Email: riko.kelter@uni-siegen.de}\\
             \emph{Draft version 1.0, 05/06/2020. This paper has not been peer reviewed. Please do not copy or cite without author's permission.} 
}


\maketitle

\begin{abstract}
Hypothesis testing is a central statistical method in psychology and the cognitive sciences. However, the problems of null hypothesis significance testing (NHST) and $p$-values have been debated widely, but few attractive alternatives exist. This article introduces the \texttt{fbst} R package, which implements the \textit{Full Bayesian Significance Test (FBST)} to test a sharp null hypothesis against its alternative via the $e$-value. The statistical theory of the FBST has been introduced by \cite{Pereira1999} more than two decades ago and since then, the FBST has shown to be a Bayesian alternative to NHST and $p$-values with both theoretical and practical highly appealing properties. The algorithm provided in the \texttt{fbst} package is applicable to any Bayesian model as long as the posterior distribution can be obtained at least numerically. The core function of the package provides the Bayesian evidence against the null hypothesis, the $e$-value. Additionally, $p$-values based on asymptotic arguments can be computed and rich visualisations for communication and interpretation of the results can be produced. Three examples of frequently used statistical procedures in the cognitive sciences are given in this paper which demonstrate how to apply the FBST in practice using the \texttt{fbst} package. Based on the success of the FBST in statistical science, the \texttt{fbst} package should be of interest to a broad range of researchers in psychology and the cognitive sciences and hopefully will encourage researchers to consider the FBST as a possible alternative when conducting hypothesis tests of a sharp null hypothesis.
\end{abstract}

\section*{Introduction}
Hypothesis testing is a widely used method in the cognitive sciences and in experimental psychology. However, the recently experienced replication crisis troubles experimental sciences and the underlying problems are still widely debated \citep{Pashler2012,Pashler2012a,Wasserstein2019,Haaf2019}. Among the identified problems is the inappropriate use and interpretation of $p$-values, which are used in combination with null hypothesis significance tests (NHST) \citep{Benjamin2019,benjaminRedefineStatisticalSignificance,Colquhoun2014,Colquhoun2017}. As a consequence, in 2016 the American Statistical Association issued a statement about the identified problems and recommended to consider alternatives to $p$-values or supplement data analysis with further measures of evidence:
\begin{quote}
    ``All these measures and approaches rely on further assumptions, but they may more directly address the size of an effect (and its associated uncertainty) or whether the hypothesis is correct.''\newline
     \cite[p.~132]{wasserstein2016}
\end{quote}
Due to the problems with NHST and $p$-values, the editors of \textit{Basic and Applied Social Psychology} even decided to ban p-values and NHST completely from their journal.

In the recent literature various proposals have been made how to improve the reproducibility of research and the quality of statistical data analysis, in particular the reliability of statistical hypothesis tests. These proposals range from stricter thresholds for stating statistical significance \citep{benjaminRedefineStatisticalSignificance} to more profound methodological changes \citep{Kruschke2018b,Wagenmakers2016,Morey2016}. In the last category, an often stated solution is a shift towards Bayesian data analysis \citep{Wagenmakers2016,Kruschke2018b,Kruschke2012,Ly2016a,Ly2016}. The advantages of such a shift include the adherence of Bayesian methods to the likelihood principle \citep{Birnbaum1962}, which has important implications. Some of them are the simplified interpretation and appealing properties of Bayesian interval estimates for quantifying the uncertainty in parameter estimates \citep{Morey2016c}. Others are given by the independence of results of the researcher's intentions \citep{Kruschke2018,Berger1988a,Edwards1963} as well as the ability to make use of optional stopping \citep{Rouder2014}. The last property is, in particular, appealing in practical research, as it allows to stop recruiting participants and report the results based on the collected data in case they already show overwhelming evidence. Notice that this is not permitted when making use of NHST and $p$-values, which can lead to financial and ethical problems, in particular in the biomedical and psychological sciences.

Considering Bayesian alternatives to NHST and $p$-values, the most prominent approach to Bayesian hypothesis testing is the Bayes factor which was invented by \cite{Jeffreys1931}, see also \cite{Etz2015}. The Bayes factor is often advocated as a Bayesian alternative to the frequentist $p$-value when it comes to hypothesis testing, in particular in the cognitive sciences and psychology \citep{VanDeSchoot2017,Wagenmakers2016,Wagenmakers2010,Ly2016,VanDoorn2019,VanDongen2019}. However, there are also other approaches like Bayesian equivalence testing based on the region of practical equivalence (ROPE) \citep{Kruschke2013,Kruschke2015,Kruschke2018,Kruschke2018a,Westlake1976,Kirkwood1981,Kelter2020BayesianPosteriorIndices,Liao2020} which are based on an analogy to frequentist equivalence tests \citep{Lakens2017,Lakens2018}. Also, there exist various other measures and alternatives to test hypotheses in the Bayesian approach, including the MAP-based $p$-value \citep{Mills2017}, the probability of direction (PD) \citep{Makowski2019,Makowski2019a} and the Full Bayesian Significance Test (FBST) \citep{Pereira1999,Stern2003,Madruga2001,Madruga2003,Pereira2008,Pereira2020,Esteves2019}. In contemporary literature, there is still a debate about which Bayesian measure to use in which setting for scientific hypothesis testing, and while some authors argue in favour of the Bayes factor \citep{Wagenmakers2016,Etz2016,Kelter2020BMCJasp}, there is also criticism about the focus on the Bayes factor in the cognitive sciences \citep{Tendeiro2019,Greenland2019}. By now, comparisons of different Bayesian posterior indices are rare, but the existing results show that it is useful to consider various different Bayesian approaches to hypothesis testing depending on the research goal and study design, see \cite{Kelter2020BayesianPosteriorIndices,Makowski2019,Liao2020}.

In this paper, attention is directed to one specific Bayesian alternative to NHST and $p$-values, the Full Bayesian Significance Test (FBST) and the $e$-value, and the R package \texttt{fbst} is introduced. The FBST has been developed over two decades ago in the statistical literature \citep{Pereira1999}, and since has been employed successfully in a broad range of scientific areas and applications. It is not possible to cover all theoretical and practical work which has been pursued concerning the FBST in the last two decades, and for a concise review, we refer the reader to \cite{Pereira2020}. The R package \texttt{fbst} introduced in this paper offers an intuitive and widely applicable software implementation of the FBST and the $e$-value. The package has been designed to work in combination with widely used R packages for fitting Bayesian models in the cognitive sciences and psychology and offers appealing visualisations to communicate and share the results of an analysis with colleagues.

The structure of this paper is as follows: First, we describe the underlying theory of the FBST and the $e$-value. Second, we give information about the available functionality and software implementation of the package. Subsequently, we demonstrate with two examples of widely used statistical models in psychological research how the FBST can be applied in practice via the \texttt{fbst} package. Finally, we conclude by drawing attention to the benefits and limitations of the package and give some ideas about future extensions. In summary, the FBST and $e$-value could be an appealing Bayesian alternative to NHST and $p$-values which has been widely under-utilised by now in the cognitive sciences and psychology. This clearly can be attributed to the dearth of accessible software implementations, one of which is presented in form of the R package introduced in this paper. The \texttt{fbst} package hopefully will foster critical discussion and reflection about different approaches to Bayesian hypothesis testing and allow to pursue further research to investigate the relationship between different posterior indices for significance and effect size \citep{Kelter2020BayesianPosteriorIndices,Makowski2019,Liao2020}.

\section*{The FBST and the $e$-value}
This section describes the statistical theory behind the FBST and the $e$-value in more detail. The philosophical basis (or conceptual approach) is first described briefly, and subsequently, the necessary notation is introduced.

\subsection*{Conceptual approach of the FBST}
The Full Bayesian Significance Test was first introduced by \cite{Pereira1999} more than two decades ago as a Bayesian alternative to traditional frequentist null hypothesis significance tests. It was invented to test a \textit{sharp} (or precise) point null hypothesis $H_0$ against its alternative $H_1$.

Traditional frequentist approaches measure the inconsistency of the observed data with a null hypothesis $H_0$ \citep{Kempthorne1976,Cox1977}. Frequentist hypothesis tests employ $p$-values to order the \textit{sample space} according to increasing inconsistency with the hypothesis. Notice, that a $p$-value is defined as the probability of obtaining a result (which, of course, is located in the sample space) equal to or more extreme than the one observed under the assumption of the null hypothesis $H_0$ \citep{Held2014}. In contrast, the $e$-value produced in the FBST aims at ordering the \textit{parameter space} according to increasing inconsistency with the observed data \citep{Pereira2008}. In formulas, traditional frequentist significance tests use the $p$-value to reject the null hypothesis $H_0$:
\begin{align*}
    p=Pr(x\in C|\theta_0)
\end{align*}
Here, $C$ often is the set of sample space values $x\in \mathcal{X}$ (where $\mathcal{X}$ is the sample space) for which a test statistic $T_{\theta_0}$ (derived under the assumption of the null hypothesis value $\theta_0$) is at least as large as the test statistic value $t$ calculated from the observed data. The set $C$ can be interpreted as the sample space values $x\in \mathcal{X}$ which are at least as \textit{inconsistent} with the null hypothesis $H_0$ as the observed data. The $p$-value now quantifies the evidence against $H_0$ by calculating the probability of sample space values $x$ being located precisely in this set \citep{Casella2002a}.

The idea put forward in \cite{Pereira1999} and \cite{Pereira2008} is simple: Instead of considering the sample space, a Bayesian should inspect the \textit{tangential set} $T$ of parameter values (which are, of course, located in the parameter space). This set consists of all parameter values which are \textit{more consistent} with the observed data $x$ than $\theta_0$, which is the Bayesian evidence $ev$. Here, $\overline{ev}$ is defined as
\begin{align*}
    \overline{ev} = Pr(\theta \in T|x)
\end{align*}
and $ev=1-\overline{ev}$. $ev$ can be interpreted as the evidence \textit{in favour} of the null hypothesis $H_0$, while $\overline{ev}$ is interpreted as the evidence \textit{against} $H_0$. This latter value is the probability of all parameter values $\theta$ which are \textit{more consistent} with the data $x$ than the null value $\theta_0$. The conceptual approach of the FBST consists, as a consequence, of constructing a duality Bayesian theory and frequentist sampling theory. This duality is constructed between frequentist significance measures which are based on ordering the \textit{sample space} according to increasing inconsistency with the data, and the Bayesian e-value, which is based on ordering the \textit{parameter space} according to increasing inconsistency with the observed data. This conceptual basis ensures that the FBST allows a seamless transition to Bayesian data analysis for researchers who are acquainted with NHST and $p$-values. The FBST produces the $e$-value which can be interpreted similarly to the frequentist $p$-value and little methodological changes are required. However, the consequences of the conceptual basis of the FBST are substantial: As the quantity $\overline{ev}$ is a fully Bayesian quantity, it allows statements in terms of probability to quantify the evidence. Traditional frequentist measures like $p$-values do not make probabilistic statements about the parameter (because they are computed over the sample space instead of the parameter space), which is questionable as the goal of the study or experiment is to quantify the uncertainty about a given research hypothesis, which naturally should be done via probability measures \citep{Howie2002,Berger1988a}. As a consequence, the FBST and the $e$-value follow the likelihood principle \citep{Birnbaum1962,Basu1975,Berger1988a}, which brings several advantages with it:
\begin{itemize}
    \item{Researchers can use optional stopping. This means that they are allowed to stop recruiting participants or even abort an experiment and readily report the results when only a fraction of the data already shows overwhelming evidence for or against the hypothesis under consideration \citep{Edwards1963,Rouder2014}.}
    \item{Censored data (which are often observed in longitudinal studies or clinical trials in the cognitive sciences and psychology) can be interpreted easily \citep{Berger1988a}. The likelihood contribution of a single observation in a study where no censoring was possible is equal to the likelihood contribution of a single observation in a study where censoring is possible but did not occur (for the single observation considered). This simplifies the analysis and interpretation of statistical models which include censoring mechanisms, see \cite[Chapter 4]{Berger1988a}.}
    \item{As highlighted by \cite{Edwards1963}, \cite{Wagenmakers2016}, and \cite{Kruschke2018a}, the result of a hypothesis test (in this case, the FBST), is not influenced by the researchers' intentions. This last property is substantial for improving the reliability of research in the cognitive sciences and psychology, see \cite{McElreath2015}.}
\end{itemize}

\subsection*{Statistical theory of the FBST}
In this section we introduce the necessary mathematical notation for a rigid understanding of the FBST. The FBST can be used with any standard parametric statistical model, where $\theta \in \Theta \subseteq \mathbb{R}^p$ is a (vector-valued) parameter of interest, $p(x|\theta)$ is the model likelihood and $p(\theta)$ is the prior distribution for the parameter $\theta$ of interest. A sharp (or expressed equivalently, precise) hypothesis $H_0$ makes a statement about the parameter $\theta$: Specifically, the null hypothesis $H_0$ states that $\theta$ lies in the so-called \textit{null set} $\Theta_{H_0}$. For simple point null hypotheses like $H_0:\theta=\theta_0$, which are often used in practice, this null set consists of the single parameter value $\theta_0$ so that the null set can be written as $\Theta_{H_0} = \{\theta_0 \}$. As detailed in the previous section, the conceptual approach of the FBST is to state the Bayesian evidence against $H_0$, the $e$-value. This value is the proposed Bayesian replacement of the traditional $p$-value. To construct the $e$-value, \cite{Pereira2008} introduced the posterior \textit{surprise function} $s(\theta)$ which is defined as follows:
\begin{align}
    s(\theta):=\frac{p(\theta|x)}{r(\theta)} 
\end{align}
The surprise function $s(\theta)$ is the ratio of the posterior distribution $p(\theta|x)$ and a suitable \textit{reference function} $r(\theta)$. The first thing to note is that two important special cases are given by a flat reference function $r(\theta)=1$ or any prior distribution $p(\theta)$ for the parameter $\theta$. First, when a flat reference function is selected the surprise function recovers the posterior distribution $p(\theta|x)$. Second, when any prior distribution is used as the reference function, one can interpret parameter values $\theta$ with a surprise function value $s(\theta)\geq 1$ as being corroborated by the observed data $x$. In contrast, parameter values $\theta$ with a surprise function $s(\theta)<1$ indicate that they have not been corroborated by the data. The next step is to calculate the supremum $s^{*}$ of the surprise function $s(\theta)$ over the null set $\Theta_{H_0}$.
\begin{align*}
    s^{*}:=s(\theta^{*})=\sup\limits_{\theta \in \Theta_{H_0}}s(\theta)
\end{align*}
This supremum is subsequently used in combination with the tangential set, which has been introduced in the last section. \cite{Pereira2008} defined
\begin{align}\label{eq:tangentialSet}
    T(\nu):=\{\theta \in \Theta|s(\theta)\leq \nu \}
\end{align}
and the tangential set $\overline{T}(\nu)$ to the sharp null hypothesis $H_0$ is then given as follows:
\begin{align}
    \overline{T}(\nu):=\Theta \setminus T(\nu)
\end{align}
When setting $\nu=s^{*}$, the tangential set $\overline{T}(\nu)$ has its unique interpretation which has been discussed in the previous section: While $T(s^{*})$ includes all parameter values $\theta$ which are smaller or equal to the supremum value $s^{*}$ of the surprise function $s(\theta)$, the tangential set $\overline{T}(s^{*})$ includes all parameter values $\theta$ which attain a \textit{larger} surprise function value than the supremum $s^{*}$ of the null set.

The final step to obtain the $e$-value, the Bayesian evidence against $H_0$, is to make use of the \textit{cumulative surprise function} $W(\nu)$
\begin{align}
    W(\nu):=\int_{T(\nu)}p(\theta|x)d\theta
\end{align}
The cumulative surprise function $W(\nu)$ is simply an integral of the posterior distribution $p(\theta|x)$ over all parameter values with surprise function values $s(\theta)\leq \nu$. Setting $\nu=s^{*}$, the cumulative surprise function $W(s^{*})$ becomes the integral of the posterior $p(\theta|x)$ over $T(s^{*})$. This is the integral of the posterior $p(\theta|x)$ over all parameter values which have a surprise function value $s(\theta)\leq s^{*}$. The $e$-value is then given as
\begin{align}\label{eq:evalue}
    \overline{\text{ev}}(H_0):=\overline{W}(s^*)
\end{align}
\begin{figure*}[!h]
\centering
\includegraphics[width=1\textwidth]{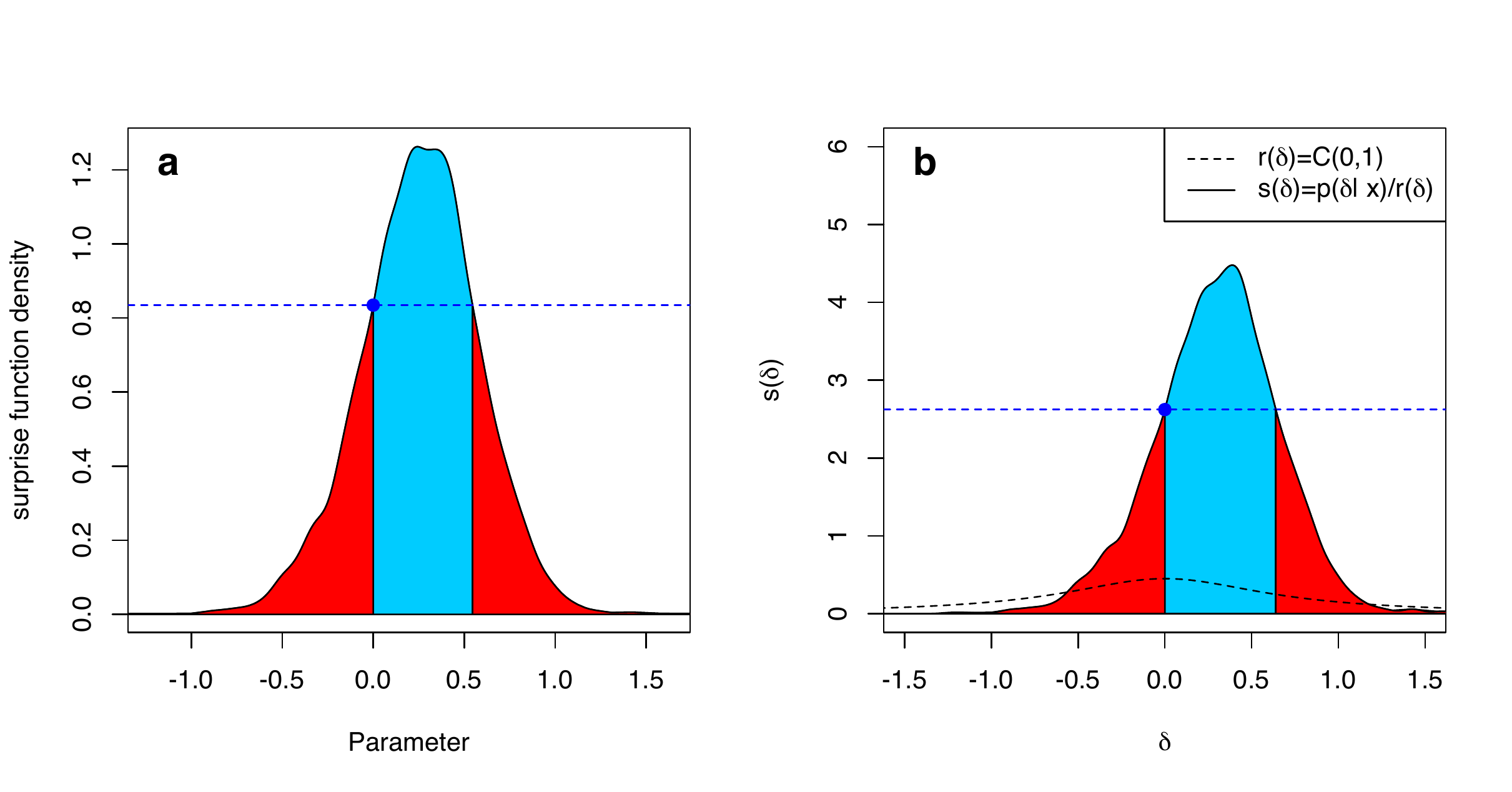}
\caption{The FBST and the $e$-value $\overline{\text{ev}}(H_0)$ against $H_0:\delta=0$ in a Bayesian two-sample t-test, where $\delta$ is the effect size. (a): A flat reference function $r(\delta)=1$ is used, and the solid line is the resulting posterior distribution $p(\delta|x)$ after observing the data. The supremum over the null set $s^{*}=0$ is visualised as the blue point. The blue shaded area corresponds to the cumulative surprise function $\overline{W}(0)$, which is the integral over the tangential set $\overline{T}(0)$ of $H_0:\delta =0$. This is the $e$-value $\overline{\text{ev}}(H_0)$ against $H_0$. The red area is the integral $W(0)$ over $T(0)$, and equals the $e$-value ev$(H_0)$ in favour of $H_0:\delta=0$. (b): The same situation as in (a), but now a Cauchy $C(0,1)$ prior has been used as reference function $r(\delta)$.}
\label{fig:fbst}
\end{figure*}
Here $\overline{W}(\nu):=1-W(\nu)$. Figure \ref{fig:fbst}a visualises the FBST and the $e$-value $\overline{\text{ev}}(H_0)$. The solid line shows the posterior distribution $p(\delta|x)$ of the effect size $\delta$ after observing the data $x$, and is produced by a Bayesian two-sample t-test \citep{Kelter2020JORSBayest}. A flat reference function $r(\delta)=1$ was selected in figure \ref{fig:fbst}a. The supremum over the null set $\Theta_{H_0}=\{0\}$ is $s^{*}=s(0)$, shown as the blue point. The horizontal blue dashed line visualises the boundary between $T(0)$ and $\overline{T}(0)$, and values with posterior density $p(\delta)>p(0)$ are located in $\overline{T}(0)$, while values with $p(\delta)\leq p(0)$ are located in $T(0)$. The blue shaded area is the cumulative surprise function $\overline{W}(0)$, which is the integral over the tangential set $\overline{T}(0)$ against $H_0:\delta =0$. This is the $e$-value $\overline{\text{ev}}(H_0)$ against $H_0$, the Bayesian evidence against the sharp null hypothesis. The red shaded area is the integral $W(0)$ over $T(0)$, which equals the $e$-value ev$(H_0)$ in favour of $H_0:\delta=0$. Figure \ref{fig:fbst}b shows the same situation, but now the reference function is selected as a wide Cauchy prior $C(0,1)$, so that the surprise function becomes
\begin{align*}
    s(\delta)=p(\delta|x)/c(\delta)
\end{align*}
where $c(\delta)$ is the p.d.f. of the $C(0,1)$ Cauchy distribution. Although the situation seems similar to figure \ref{fig:fbst}a, the scaling on the $y$-axis now is different. Also, the evidence has changed based on the new surprise function and the interpretation of the surprise function has changed, too. While in figure \ref{fig:fbst}a, the surprise function could be interpreted as the posterior distribution, now it is interpreted as follows: If one assumes a Cauchy prior $C(0,1)$ on the effect size $\delta$, then parameters with a surprise function value $s(\delta)\geq 1$ can be interpreted as being corroborated by the data. Parameter values with a surprise function $s(\delta)<1$ are interpreted as not being corroborated by the data.

\cite{Pereira1999} formally defined the $e$-value ev$(H_0)$ in \textit{support} of $H_0$ as 
\begin{align}
    \text{ev}(H_0):=1-\overline{\text{ev}}(H_0)
\end{align}
but notice that one can not interpret this value as evidence \textit{against} $H_1$. This can be attributed to the fact that $H_1$ is not even a sharp hypothesis, see Definition 2.2 in \cite{Pereira2008}.

It is crucial to note that it is not possible to utilise the $e$-value ev$(H_0)$ to \textit{confirm} the null hypothesis $H_0$ \citep{Kelter2020BayesianPosteriorIndices}. However, the FBST can be generalized in to an extended framework which then allows for hypothesis confirmation and itself is an active topic of ongoing research \citep{Esteves2019}. Additionally, the $e$-value ev$(H_0)$ can be used to reject $H_0$ if ev$(H_0)$ is sufficiently small based on asymptotic arguments \citep[Section 5]{Pereira2008}. \cite{Pereira2008} showed that the distribution of the $e$-value is a Chi-square distribution
\begin{align}\label{eq:evPVal}
    \text{ev}(H_0) \sim \chi_k^2(||m-M||^2)    
\end{align}
where $M$ is the posterior mode calculated over the entire parameter space $\Theta$ and $m$ is the posterior maximum over $\Theta_{H_0}$. The $p$-value associated with the Bayesian evidence in support of $H_0$ is then calculated as the superior tail of the $\chi^2$ density with $k-h$ degrees of freedom, starting from $-2\lambda(m_0)$. Here, $k$ is the dimension of the parameter space $\Theta$ and $h$ is the dimension of the null set $\Theta_{H_0}$. The quantity $m_0$ is the observed value and $\lambda(t)=\ln l(t)$ is the logarithm of the relative likelihood function, where $l(t)=L(t)/L(M)$ is the relative likelihood. Denoting $F_{k-h}$ as the Chi-square distribution's cumulative distribution function with $k-h$ degrees of freedom ($F_k$ analogue), the $p$-value associated with the Bayesian $e$-value $\text{ev}(H_0)$ is then computed as
\begin{align}
    pv_{0}=1-F_{k-h}(-2\lambda(m_0))    
\end{align}
This latter $p$-value has a frequentist interpretation. The $p$-value based on equation (\ref{eq:evPVal}) can be expressed as
\begin{align}\label{eq:evPVal2}
    ev_0 = F_k(||m_0-M_0||^2)    
\end{align}
and can be interpreted as a Bayesian significance value which quantifies the probability of obtaining $\text{ev}(H_0)$ or even \textit{less} evidence in support of the null hypothesis $H_0$. Consequently, after observing $m_0$ and $M_0$ one only needs to calculate the euclidian distance $d_0=||m_0-M_0||^2$ and the value of the $\chi_{k}^2$ distribution's cumulative distribution function of this distance is the corresponding $p$-value. Based on a threshold (like $0.05$) one can decide to reject the null hypothesis $H_0:\theta=\theta_0$ or not.

However, if a $p$-value is required which is closest to the frequentist $p$-value in interpretation, one should use the standardized $e$-value $\text{sev}(H_0)$, as defined in \cite[Section 2.2]{Borges2007} and in \cite[Section 3.3]{Pereira2020}.  The standardized $e$-value is defined as:
\begin{align*}
    \overline{\text{sev}}(H_0)=F_{k-h}(F^{-1}_{k}(\overline{\text{ev}}))
\end{align*}
Here, $F^{-1}_{k}$ is the quantile function of the cumulative distribution function of the $\chi_{k}^2$ distribution with $k$ degrees of freedom. $\overline{\text{sev}}(H_0)$ can, as a consequence, be interpreted as the probability of obtaining less evidence than $\overline{\text{ev}}(H_0)$ against the null hypothesis $H_0$. Defining
\begin{align*}
    \text{sev}(H_0)=1-\overline{\text{sev}}(H_0)
\end{align*}
$\text{sev}(H_0)$ can then be interpreted as the probability of obtaining $\overline{\text{ev}}(H_0)$ or more evidence against $H_0$, which is closely related to the interpretation of a frequentist $p$-value. However, the $p$-value operates in the sample space while the standardized $e$-value operates in the parameter space. The standardized $e$-value can be used as a Bayesian replacement of the frequentist $p$-value, while being very similar in interpretation. For theoretical properties of $\text{sev}(H_0)$ see \cite{Borges2007} and \cite{Pereira2020}.

In the examples below, the Bayesian evidence against $H_0$, the $e$-value $\overline{\text{ev}}(H_0)$ is reported and also the standardized $e$-values $\text{sev}(H_0)$ are given.

\section*{Overview and functionality of the fbst package}
The centerpiece of the \texttt{fbst} package is the \texttt{fbst()} function, which is used to perform the FBST. In addition to the \texttt{fbst()} function, the package provides customised \texttt{summary()} and \texttt{plot()} functions which allow users to print the results of a FBST or obtain a visualisation of their results to communicate and share the results. The \texttt{fbst()} function has the following structure:
\begin{lstlisting}[caption={The \texttt{fbst()} function}\label{r:fbstFunction},language=R]
fbst(posteriorDensityDraws, nullHypothesisValue, FUN, par, 
     dimensionTheta, dimensionNullset)    
\end{lstlisting}
Here, \texttt{posteriorDensityDraws} needs to be a numeric vector holding the posterior parameter draws obtained via MCMC or any other numerical method of choice.\footnote{If the posterior is available in closed form, one can directly sample from it and provide the argument with the samples.} The argument \texttt{nullHypothesisValue} is the value specified in the null hypothesis $H_0:\theta=\theta_0$, and \texttt{dimensionTheta} is the dimension of the parameter space $\Theta$. \texttt{dimensionNullset} is the dimension of the null set $\Theta_{H_0}$, and \texttt{FUN} and \texttt{par} are additional arguments which only need to be specified when a user-defined reference function $r(\theta)$ is desired. In general, \texttt{FUN} should be the name of the reference function which should be used and \texttt{par} should be a list of parameters which this reference function utilises (e.g. the location and scale parameters when the reference function is a Cauchy prior). Details will be given in the examples below.

The \texttt{fbst()} function returns an object of the class \texttt{fbst}, which stores several useful details and the results of the conducted FBST. To obtain a concise summary of the FBST, the \texttt{summary()} function of the class \texttt{fbst} can be used. To visualise the FBST, the \texttt{plot()} function of the \texttt{fbst} class can be used. Details are provided in the examples below.

From an algorithmic perspective, the \texttt{fbst} package proceeds via the following steps when computing the e-value via the \texttt{fbst()} function:
\begin{enumerate}
    \item{Based on the posterior parameter samples \\\texttt{posteriorDensityDraws}, the posterior density $p(\theta|x)$ is estimated via a Gaussian kernel density estimator, resulting in a posterior density estimate $\hat{p}(\theta|x)$. The Gaussian kernel is used due to well-known Bayesian asymptotics of posterior distributions, the Bernstein-von-Mises theorem \citep{Held2014}.}
    \item{Based on this posterior density estimate $\hat{p}(\theta|x)$, the surprise function $s(\theta)$ is estimated (i) as the posterior density estimate $\hat{p}(\theta|x)$ if no arguments \texttt{FUN} and \texttt{par} are supplied so that a flat reference function $r(\theta)=1$ is used as default, or (ii) as the ratio $\hat{p}(\theta|x)/r(\theta)$ if arguments \texttt{FUN} and \texttt{par} are supplied. The result is a surprise function estimate $\hat{s}(\theta)$.}
    \item{The surprise function estimate $\hat{s}(\theta)$ is evaluated at the null hypothesis value supplied via the argument \texttt{nullHypothesisValue}, resulting in the value $\hat{s}_0$.}
    \item{The $e$-value $\overline{\text{ev}}(H_0)$ is computed via numerical integration of the posterior density estimate $\hat{p}(\theta|x)$ over the tangential set $\overline{T}(H_0)$, which is determined via a linear search algorithm on the vector \texttt{posteriorDensityDraws} by including all values $\theta$ which fulfill the condition $\hat{s}(\theta)>\hat{s}_0$.}
    \item{The $p$-value associated with the $e$-value $\text{ev}(H_0)$ in favour of the null hypothesis $H_0$ and the standardized $e$-values sev$(H_0)$ are computed.} 
\end{enumerate}
In summary, the FBST is based only on simple numerical optimization and integration which makes it a computationally cheap option. This is a benefit, in particular, when the parameter space $\Theta$ is high-dimensional \cite{Pereira2020,Stern2003,Kelter2020BayesianPosteriorIndices}. Also, the presence of nuisance parameters does not trouble the computation unlike it is the case for example for the Bayes factor, as computing the marginal likelihoods can quickly become difficult then \citep{Stern2003}.

\section*{Example 1: Two-sample Bayesian t-test}
As a preliminary note, all analyses can be reproduced by following the provided code.\footnote{However, a supplementary replication codebook is provided at the Open Science Foundation under \url{https://osf.io/u6xnc/}.}. To demonstrate how to use the \texttt{fbst} package, we start with the two-sample t-test, a widely used statistical model in the cognitive sciences \citep{Nuijten2016}. We use the two-sample Bayesian t-test of \cite{Rouder2009} together with simulated data. The recommended medium Cauchy prior $C(0,\sqrt{2}/2)$ was assigned to the effect size $\delta$. Observations in the first group were simulated as $\mathcal{N}(0,1.7)$, and observations belonging to the second group were generated according to the $\mathcal{N}(0.8,3)$ distribution. As a consequence, the resulting true effect size $\delta$ according to \cite{cohen_statistical_1988} is given as
\begin{align*}
    \delta = \frac{0-0.8}{\sqrt{(1.7^2+3^2)/2}} \approx -0.33
\end{align*}
which equals a small effect size. The code to simulate the data is given in listing \ref{r:listing1}.

\begin{figure}[!h]
\centering
\includegraphics[width=0.5\textwidth]{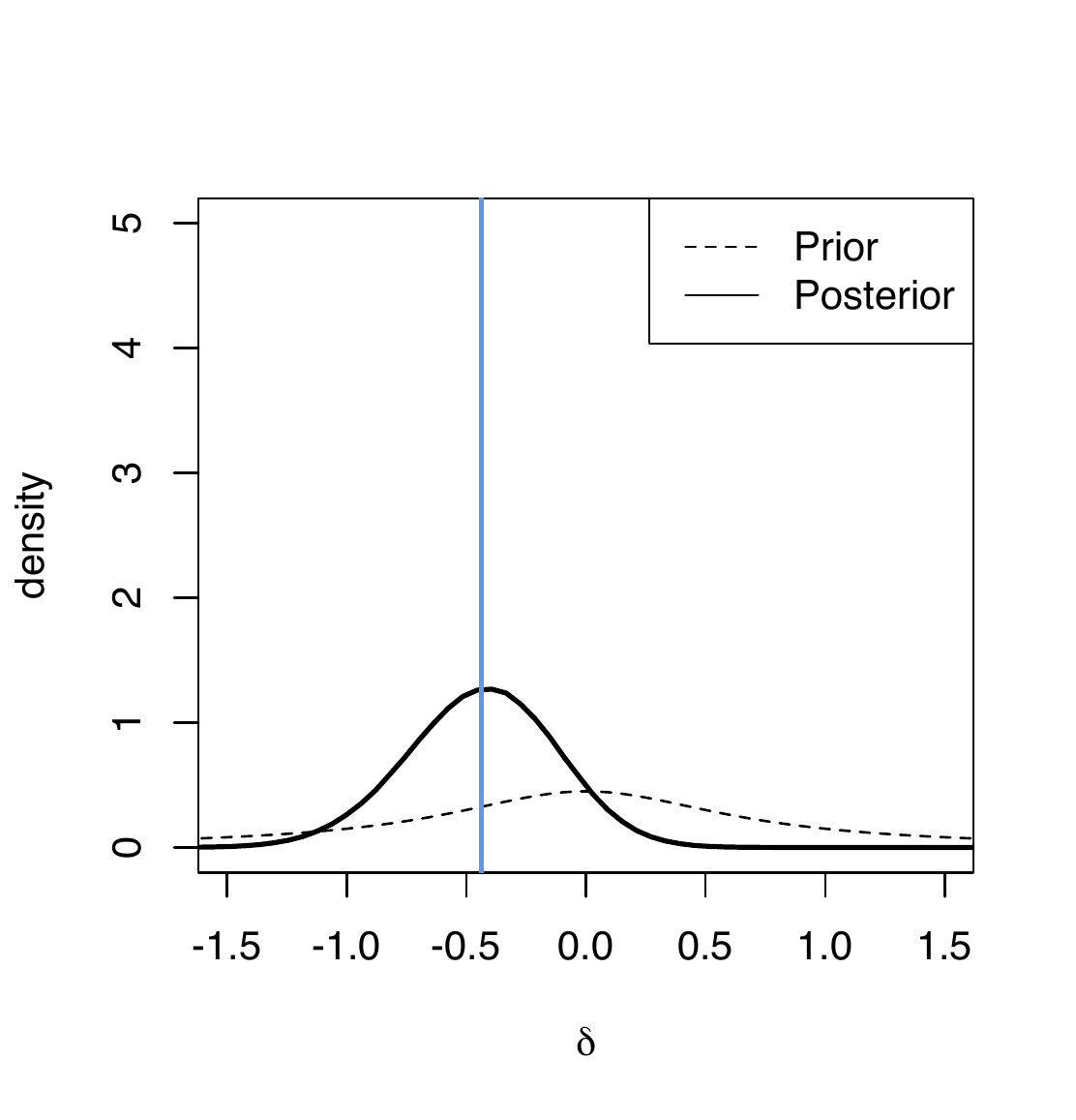}
\caption{Prior-posterior plot for Example 1}
\label{fig:example1priorposterior}
\end{figure}

\begin{lstlisting}[caption={Example 1 - Simulation of data}\label{r:listing1},language=R]
    set.seed(57)
    grp1=rnorm(18,0,1.7)
    grp2=rnorm(18,0.8,3)
\end{lstlisting}
The corresponding Bayes factor $BF_{10}$ for the alternative hypothesis $H_1:\delta \neq 0$ against the null hypothesis $H_0:\delta = 0$ is given as $BF_{10}=0.91$, which does not indicate evidence worth mentioning according to \cite{jeffreys1961} or \cite{VanDoorn2019}. The slight favour towards $H_0$ can be attributed to the medium Cauchy prior used, which centres the prior probability mass closely around small effect sizes (and no effect, too). Figure \ref{fig:example1priorposterior} shows a prior-posterior plot for the example. The code to compute the Bayes factor is given in listing \ref{r:listing2}.
\begin{lstlisting}[caption={Example 1 - Producing the posterior distribution of the effect size and hypothesis testing via the Bayes factor}\label{r:listing2},language=R]
install.packages("BayesFactor")
library(BayesFactor)
p = BayesFactor::ttestBF(x=grp1,y=grp2, posterior = TRUE, iterations = 3000000, rscale = "medium")
p = as.vector(p[,4])

BF_10 = BayesFactor::ttestBF(x=grp1,y=grp2, posterior = FALSE, 
    rscale = "medium", paired = FALSE)
BF_10
\end{lstlisting}

To perform the FBST and compute the $e$-value, we first install and load the R package from CRAN by executing the code in listing \ref{r:listing3}.
\begin{lstlisting}[caption={Example 1 - Hypothesis testing via the Bayes factor}\label{r:listing3},language=R]
    install.packages("fbst")
    library(fbst)
    resFlatSim = fbst(posteriorDensityDraws = p, nullHypothesisValue = 0, dimensionTheta = 3, dimensionNullset = 2)
\end{lstlisting}
Note that in the example, the parameter space $\Theta$ consists of three parameters: The mean $\mu_1$ in the first group, the mean $\mu_2$ in the second group, and the standard deviation $\sigma^2$. As a consequence, the argument \texttt{dimensionTheta} is therefore set to \texttt{dimensionTheta=3}. The null set $\Theta_{H_0}$ consists of the set $\{\mu_1=\mu_2,\sigma^2\}$, which is two-dimensional so that \texttt{dimensionNullset = 2}. The object stored in the variable \texttt{resFlatSim} is an object of the class \texttt{fbst}, which stores several values used in the \texttt{summary()} and \texttt{plot()} functions of the package. These are available to communicate and visualise the results of the FBST. For example, we can access the $e$-value $\overline{\text{ev}}(H_0)$ as follows (see listing \ref{r:listing3b}):
\begin{lstlisting}[caption={Example 1 - The Bayesian e-value against the null hypothesis of no effect}\label{r:listing3b},language=R]
    resFlatSim@eValue
    [1] 0.8305998
\end{lstlisting}
Instead of accessing each attribute manually, to obtain a summary of the FBST and print the relevant quantities the \texttt{summary()} function of the \texttt{fbst} package provides a more convenient option:
\begin{lstlisting}[caption={Example 1 - Using the \texttt{summary()} function on an \texttt{fbst} object}\label{r:listing4},language=R]
    summary(resFlatSim)
    
    Full Bayesian Significance Test for testing a sharp hypothesis against its alternative:
    Reference function: Flat 
    Testing Hypothesis H_0:Parameter= 0 against its alternative H_1
    Bayesian e-value against H_0: 0.8305998 
    p-value associated with the Bayesian e-value in favour of the null hypothesis: 0.1461029 
    Standardized e-value: 0.0248695
\end{lstlisting}
Based on the results, we can see that there is some evidence against the null hypothesis according to the Bayesian $e$-value $\overline{\text{ev}}(H_0)$ against $H_0$ (compare equation (\ref{eq:evalue})). The corresponding $p$-value $ev_0 \approx 0.146$ is not significant if a standard threshold of $0.05$ is used, but the standardized $e$-value $\text{sev}(H_0) \approx 0.025 <0.05$ is. Note that when a $p$-value is used for hypothesis testing, it is recommended to use the standardized $e$-value \cite{Borges2007,Pereira2020}, so one would reject the null hypothesis $H_0:\delta = 0$ in this case. However, it is also possible to use only the Bayesian evidence $\overline{\text{ev}}(H_0)$ against $H_0$ without any $p$-value to quantify the evidence continuously.

\begin{figure*}[!h]
\centering
\includegraphics[width=1\textwidth]{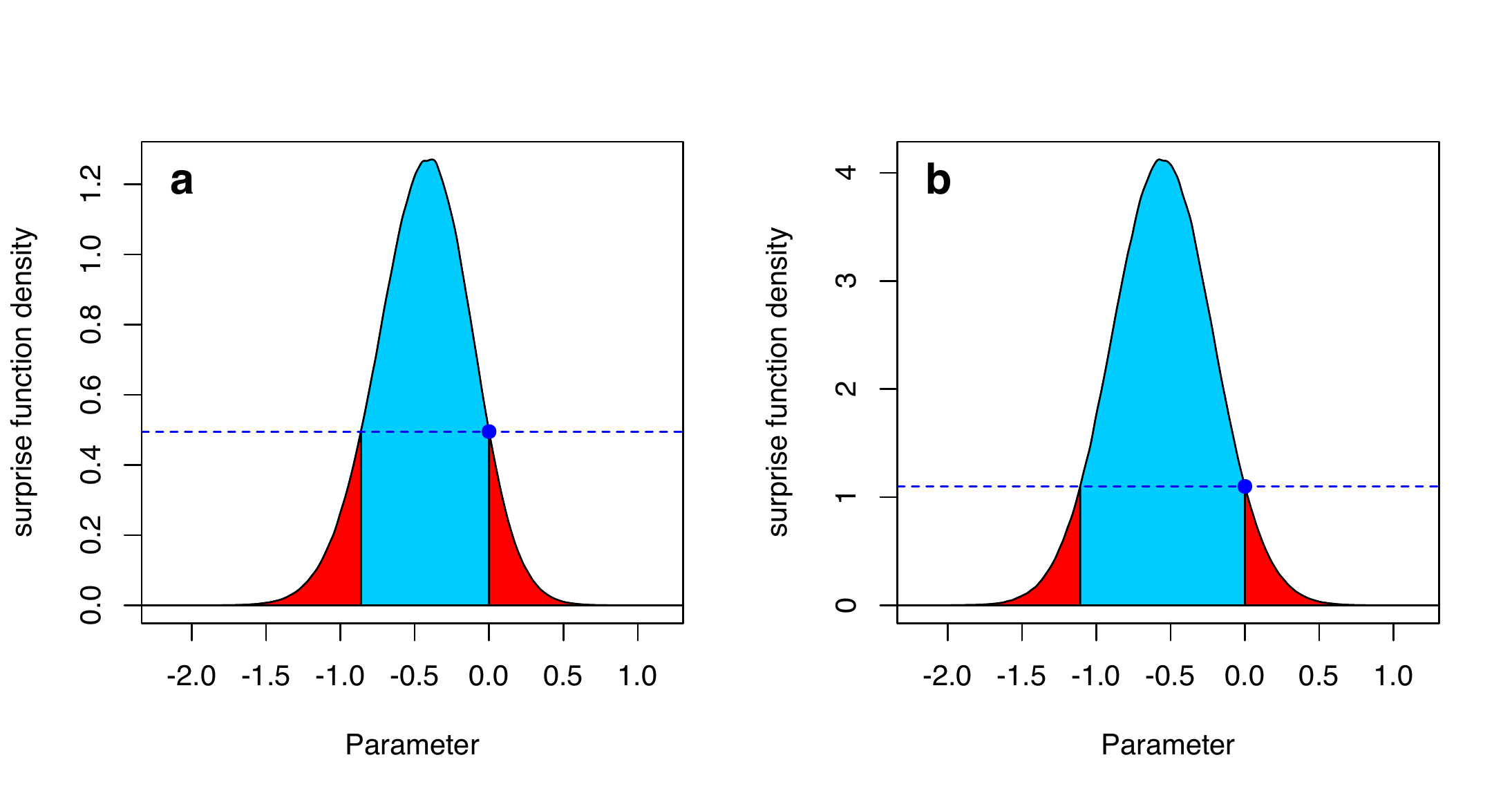}
\caption{(a) Visualisation of the FBST for the Bayesian two-sample t-test in Example 1 using a flat reference function $r(\delta)=1$; (b) Visualisation of the FBST for the Bayesian two-sample t-test in Example 1 using a medium Cauchy prior as reference function $r(\delta)=C(0,\sqrt{2}/2)$}
\label{fig:example1fbst}
\end{figure*}

To visualise the results, we use the \texttt{plot()} function of the \texttt{fbst} package:
\begin{lstlisting}[caption={Example 1 - Hypothesis testing via the Bayes factor}\label{r:listing5},language=R]
    plot(resFlatSim)
\end{lstlisting}
The result is shown in figure \ref{fig:example1fbst}a: The blue shaded area under the surprise function (which is by default the posterior distribution, that is, a flat reference function $r(\delta)=1$ is used by default by the \texttt{fbst()} function) is the Bayesian evidence against $H_0$, the $e$-value $\overline{\text{ev}}(H_0)\approx 0.83$ (compare listing \ref{r:listing4}). The red shaded area is the $e$-value $\text{ev}(H_0)$ in favour of $H_0$, which is $ev(H_0) \approx 1-0.83=0.17$.
Instead of a flat reference function $r(\delta)=1$, one could also use a more reasonable prior distribution. For example, as small to medium effect sizes are to be expected in the cognitive sciences and psychology, \cite{Rouder2009} recommended to use a medium Cauchy prior $C(0,\sqrt{2}/2)$ as a default prior on the effect size. To see if parameter values $\delta$ have been corroborated (compared to this prior assumption) by observing the data, we can use this prior as the reference function $r(\delta)=C(0,\sqrt{2}/2)$, and the resulting surprise function is shown in figure \ref{fig:example1fbst}b. The code to produce the FBST based on a Cauchy reference density is given in listing \ref{r:listing6}:
\begin{lstlisting}[caption={Example 1 - The FBST using a medium Cauchy prior as reference function}\label{r:listing6},language=R]
    resMediumSim = fbst(posteriorDensityDraws = p, nullHypothesisValue = 0, dimensionTheta = 3, dimensionNullset = 2, FUN=dcauchy, par = list(location = 0, scale = sqrt(2)/2))
    
    summary(resMediumSim)
    Full Bayesian Significance Test for testing a sharp hypothesis against its alternative:
    Reference function: User-defined 
    Testing Hypothesis H_0:Parameter= 0 against its alternative H_1
    Bayesian e-value against H_0: 0.9032063 
    p-value associated with the Bayesian e-value in favour of the null hypothesis: 0.1461029 
    Standardized e-value: 0.01189972  
\end{lstlisting}
There, the \texttt{FUN} argument is supplied with the name of the density to be used and the \texttt{par} argument is supplied with a list of arguments for this density. As the Cauchy distribution has a \texttt{location} and \texttt{scale} parameter, we supply these here. Notice that the blue point which indicates the surprise function value $s(0)$ of the null hypothesis parameter $\delta=0$ is larger than one. This means that the null hypothesis value has been corroborated by the data. However, all parameter values in the tangential set have been corroborated \textit{even more} by the data than the null value $\delta=0$.

Based on the continuous quantification, there is again strong evidence against the null hypothesis when changing the reference function to a medium Cauchy prior: More than 90\% of the posterior distribution's parameter values attain a larger surprise function value than the null hypothesis value. The resulting standardized $e$-value $\text{sev}(H_0)$ is also significant.

\section*{Example 2: Directional two-sample Bayesian t-test}
Example 1 showed how to apply the FBST in the setting of the Bayesian two-sample t-test. Example 2 is a slight modification of Example 1. Instead of testing a two-sided hypothesis, we now turn to directional hypotheses and show how these can easily be tested via the \texttt{fbst} package, too. We use data of \cite{Moore2012}, which provides the reading performance of two groups of pupils: One control group and a treatment group which was given directed reading activities. The data are freely available in the built-in data library of the open-source software JASP\footnote{See \url{www.jasp-stats.org}}. We test the hypothesis $H_0:\delta < 0$, which is equivalent to the hypothesis $H_0:\mu_1 < \mu_2$, where the measured quantity is the performance of pupils in the degree of reading power test (DRP) \citep{Moore2012}.

\begin{figure*}[!h]
\centering
\includegraphics[width=1\textwidth]{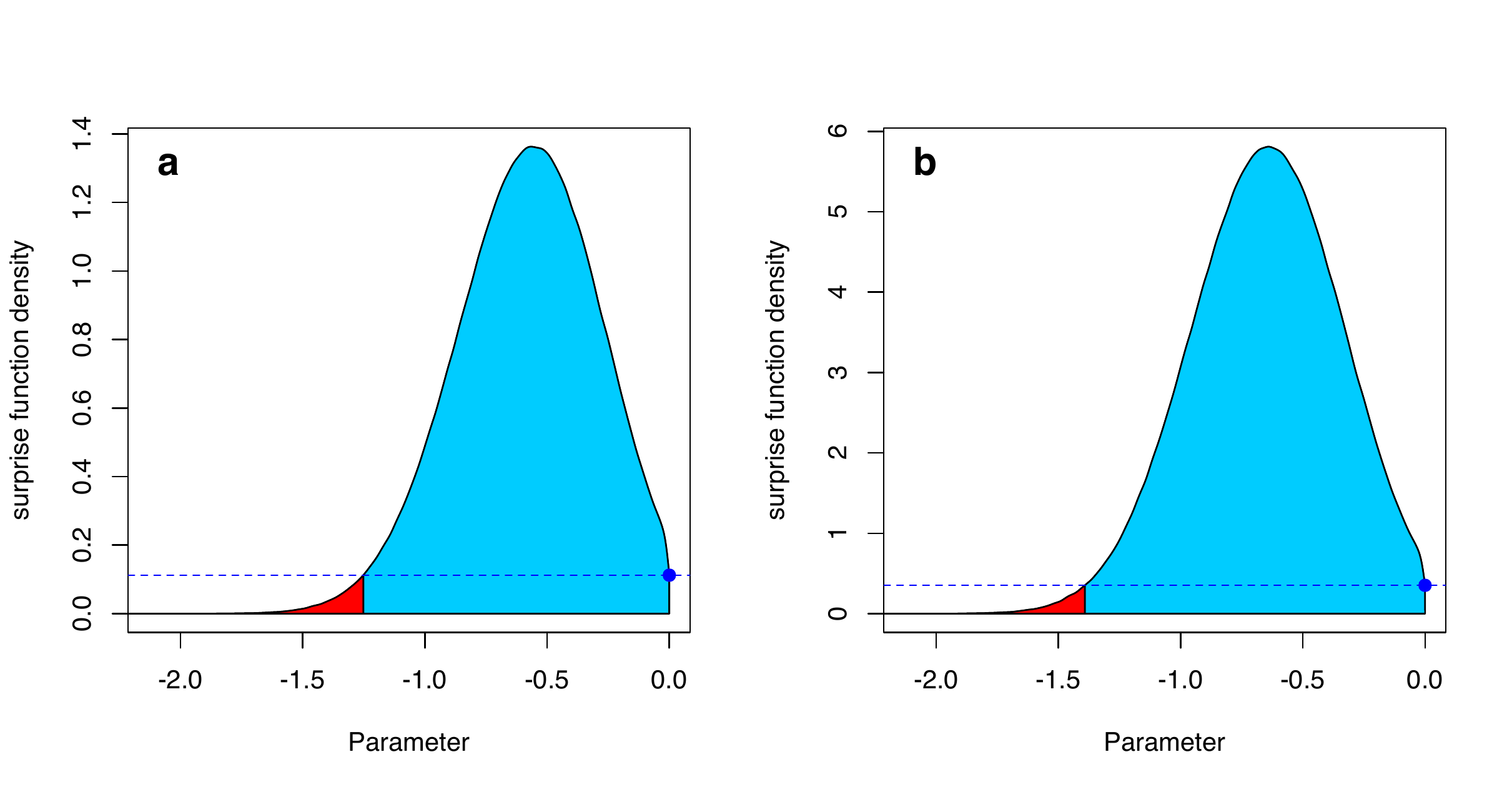}
\caption{(a) Visualisation of the FBST in Example 2 for the Bayesian two-sample t-test for testing $H_0:\delta=0$ against the one-sided hypothesis $H_1:\delta <0$, using a flat reference function $r(\delta)=1$; (b) Visualisation of the FBST in Example 2 for the Bayesian two-sample t-test for testing $H_0:\delta=0$ against the one-sided hypothesis $H_1:\delta <0$, using a wide Cauchy prior reference function $r(\delta)=C(0,1)$}
\label{fig:example2fbst}
\end{figure*}

First, we save the data in a .csv-file (which is called \texttt{DirectedReadingActivities.csv} in listing \ref{r:listing7}), set the working directory and load the data\footnote{The data set is also provided as a .csv-file at the OSF repository \url{https://osf.io/u6xnc/}.}:
\begin{lstlisting}[caption={Example 2 - Loading the data}\label{r:listing7},language=R]
setwd(' ... ') # Change to where the data are stored on your machine
library(dplyr)
dra=read.csv("DirectedReadingActivities.csv",sep=",")
head(dra)

id group g drp
1  1 Treat 0  24
2  2 Treat 0  56
3  3 Treat 0  43
4  4 Treat 0  59
5  5 Treat 0  58
6  6 Treat 0  52

treat = (dra %>% filter(group=="Treat") %>% select(drp))$drp
control = (dra %>% filter(group=="Control") %>% select(drp))$drp
\end{lstlisting}
The code to perform a standard hypothesis test based on the Bayes factor is given in listing \ref{r:listing8}, which results in $BF_{10}=4.32$, indicating moderate evidence for the alternative $H_1:\delta <0$ according to \cite{VanDoorn2019}.
\begin{lstlisting}[caption={Example 2 - Obtaining the posterior distribution of the effect size and standard Bayesian hypothesis testing via the Bayes factor}\label{r:listing8},language=R]
library(BayesFactor)
# BF closed-form of Rouder et al. (2009)
p = BayesFactor::ttestBF(x=control,y=treat, posterior = TRUE, rscale = "medium", paired = FALSE, nullInterval = c(-Inf,0), iterations = 3000000)
p = as.vector(p[,4])

BF_10 = BayesFactor::ttestBF(x=control,y=treat, posterior = FALSE, rscale = "medium", paired = FALSE, nullInterval = c(-Inf,0))
BF_10[1]

Bayes factor analysis
--------------
[1] Alt., r=0.707 -Inf<d<0 : 4.327919 ±0%

Against denominator:
  Null, mu1-mu2 = 0 
---
Bayes factor type: BFindepSample, JZS
\end{lstlisting}
The code to perform the FBST with a flat reference function $r(\delta)=1$ is given in listing \ref{r:listing9}:
\begin{lstlisting}[caption={Example 2 - Performing the FBST}\label{r:listing9},language=R]
library(fbst)
resFlatDRA = fbst(posteriorDensityDraws = p, nullHypothesisValue = 0, dimensionTheta = 3, dimensionNullset = 2)
summary(resFlatDRA)

Full Bayesian Significance Test for testing a sharp hypothesis against its alternative:
Reference function: Flat 
Testing Hypothesis H_0:Parameter= 0 against its alternative H_1
Bayesian e-value against H_0: 0.9859827 
p-value associated with the Bayesian e-value in favour of the null hypothesis: 0.06689926 
Standardized e-value: 0.001123303 
\end{lstlisting}
The dimensions of $\Theta$ and $\Theta_{H_0}$ are identical to Example 1, and the Bayesian $e$-value $\overline{\text{ev}}(H_0) \approx 0.986$ expresses strong evidence against the null hypothesis $H_0:\delta=0$. Also, the standardized $e$-value $\text{sev}(H_0) \approx 0.001 < 0.05$ is significant and leads to the same conclusion if a threshold of $0.05$ is applied. The results are visualised in figure \ref{fig:example2fbst}. Figure \ref{fig:example2fbst}a shows the FBST when a wide half-Cauchy prior $C_{+}(0,1)$ is used as the reference function $r(\delta)$ \citep{Rouder2009}\footnote{A \textit{left}-half Cauchy prior is used, as under $H_1:\delta <0$, so a priori only negative effect sizes are assumed under the alternative hypothesis.}. Figure \ref{fig:example2fbst}a is produced by the code in listing \ref{r:listing10}, where the additional parameter \texttt{rightBoundary = 0} needs to be added to inform the \texttt{plot()} function that a one-sided hypothesis was used. Should the alternative be $H_1:\delta >0$, one would supply the argument \texttt{leftBoundary = 0} to the \texttt{plot()} function.
\begin{lstlisting}[caption={Example 2 - Visualising the FBST}\label{r:listing10},language=R]
plot(resFlatDRA, rightBoundary = 0)
\end{lstlisting}
Based on the continuous quantification of evidence against $H_0$ in form of $\overline{\text{ev}}(H_0)$ and the standardized $e$-value $\text{sev}(H_0)$ one would reject the null hypothesis $H_0:\delta=0$ in favour of the alternative $H_1:\delta <0$. That is, the performance in the treatment group is better than in the control group which was not given directed reading activities.

\section*{Example 3: Bayesian logistic regression}
As a third example, we demonstrate how to use the FBST via the \texttt{fbst} package in the context of the Bayesian logistic regression model \citep{McElreath2020}. Notice that while we focus on the standard logistic model here, the procedure is applicable to any regression model of interest like probit or linear regression models. We use data from the Western Collaborative Group Study (WCGS) of \cite{Rosenman1975}, in which $3154$ healthy young men aged $39-59$ from the San Francisco area were assessed for their personality type. All were free from coronary heart disease at the start of the research. Eight and a half years later change in this situation was recorded. We use a subset of $n=3140$ participants, where 14 participants have been excluded because of incomplete data. The data set is freely available in the \texttt{faraway} R package, so we first load and prepare the data as shown in listing \ref{r:listing11}.
\begin{lstlisting}[caption={Example 3 - Loading the data}\label{r:listing11},language=R]
library(faraway)
data(wcgs)
wcgs = wcgs[complete.cases(wcgs), ]
\end{lstlisting}
For illustration purposes, we use a Bayesian logistic regression model which studies the influence of the covariates \texttt{age}, \texttt{height}, \texttt{weight}, systolic blood pressure (\texttt{sdp}), diastolic blood pressure (\texttt{dbp}), fasting serum cholesterol (\texttt{chol}) and the number of cigarettes smoked per day (\texttt{cigs}) on the outcome chronic heart disease (yes / no, variable \texttt{chd}) stored in the response variable \texttt{chd}.

\begin{figure*}[!h]
\centering
\includegraphics[width=1\textwidth]{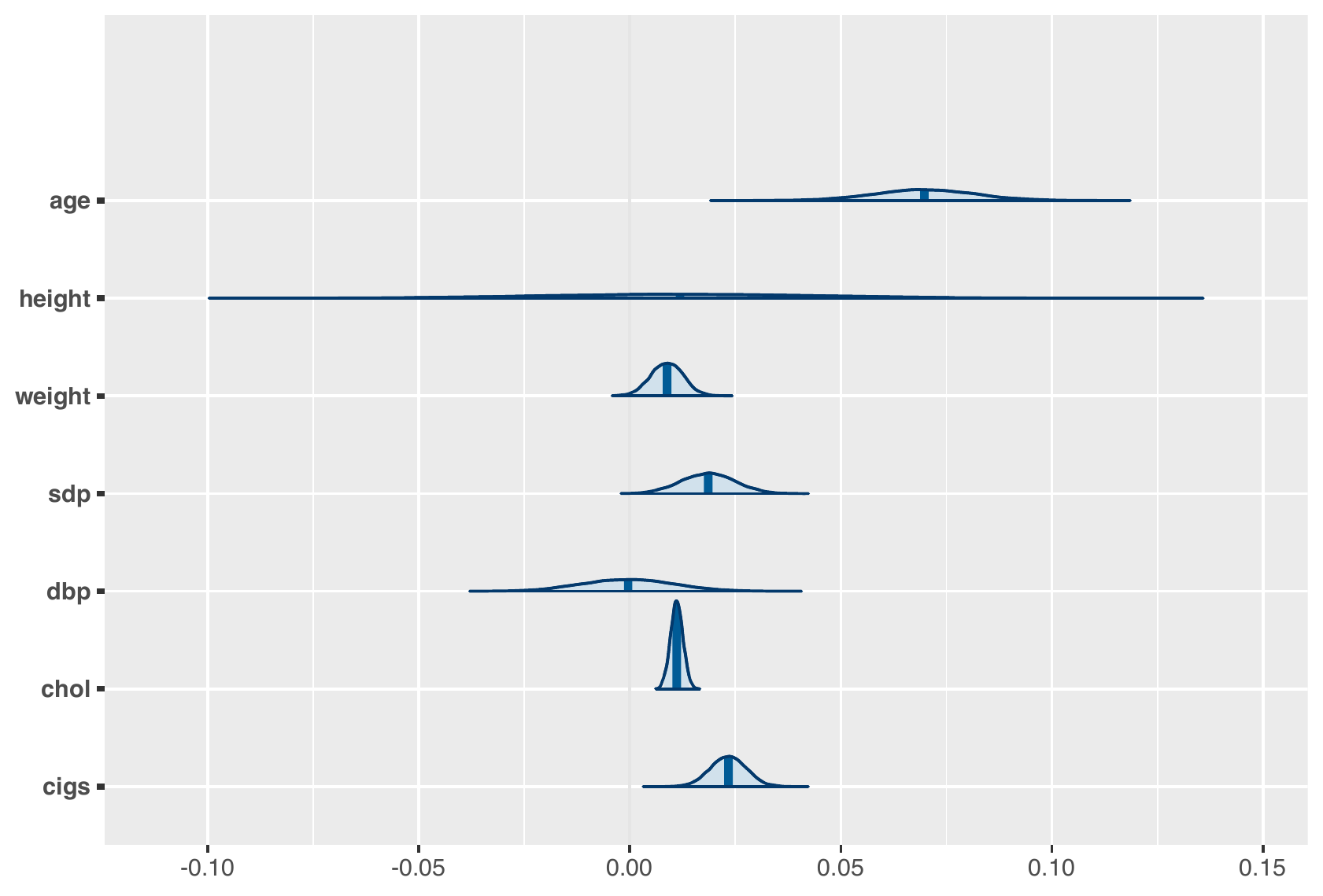}
\caption{Marginal posterior distributions of the regression coefficients $\beta_j$ in the Bayesian logistic regression model in Example 3}
\label{fig:example3}
\end{figure*}

\begin{figure*}[!h]
\centering
\includegraphics[width=1\textwidth]{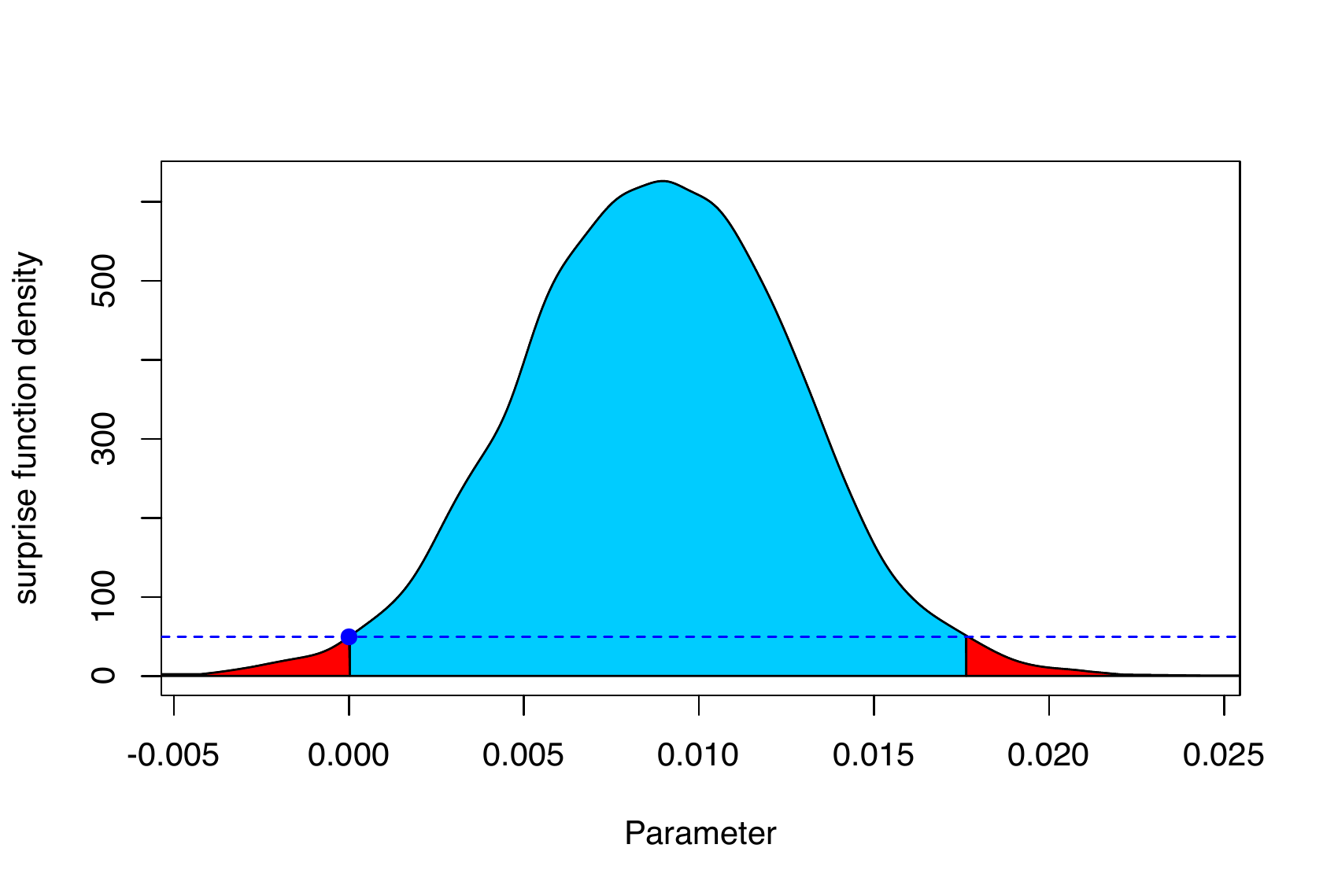}
\caption{Visualisation of the FBST for $H_0:\beta_j=0$ against $H_1:\beta_j \neq 0$ for the regression coefficient of the covariate weight in the Bayesian logistic regression model for the WCGS study}
\label{fig:example3fbst}
\end{figure*}

The model is fit via the Hamiltonian Monte Carlo sampler Stan \citep{Carpenter2017,Kelter2020} which uses the No-U-Turn sampler of \cite{Hoffman2014} to sample from the posterior distribution. We obtain the posterior distribution of the intercept and the seven regression coefficients $\beta_1,...,\beta_7$, belonging to the six covariates included in the model. The default weakly informative $\sigma \sim \exp(1)$ prior is assigned to the standard deviation $\sigma$, see \cite{Gabry2020RstanarmPriorsVignette}. We use the \texttt{rstanarm} package \citep{Goodrich2020} for fitting the Bayesian logistic regression model, and the code to prepare the data for Stan is given in listing \ref{r:listing12}.
\begin{lstlisting}[caption={Example 3 - Preparing the data for Stan}\label{r:listing12},language=R]
f1 = as.formula(paste('chd ~ age + height + weight + sdp + dbp + chol + cigs')) 
X1 <- model.matrix(f1, wcgs) # build model matrix
standata_m1 <- list(y = as.numeric(wcgs$chd)-1, X = X1, N = nrow(X1), P = ncol(X1)) # format data as list for Stan
stan_df1 <- as.data.frame(standata_m1)
\end{lstlisting}
The standard weakly informative prior distribution $\beta_j \sim \mathcal{N}(0,2.5)$ is assigned to the regression coefficients $\beta_j, j=1,...,7$, and the intercept $\beta_0$ is assigned the weakly informative default prior $\beta_0 \sim \mathcal{N}(0,10)$ recommended by \cite{Gabry2020RstanarmPriorsVignette}. Listing \ref{r:listing13} shows the code to fit the model via the \texttt{rstanarm} package, summarise and plot the results.
\begin{lstlisting}[caption={Example 3 - Fitting the Bayesian logistic regression model via Stan}\label{r:listing13},language=R]
library(rstanarm)
post_m1 <- stan_glm(f1, data = wcgs,
    family = binomial(link = "logit"), 
    prior = normal(0,2.5), 
    prior_intercept = normal(0,10), 
    QR=TRUE,
    iter = 4000,
    seed = 4711)
                  
summary(post_m1)

Model Info:
 function:     stan_glm
 family:       binomial [logit]
 formula:      chd ~ age + height + weight + sdp + dbp + chol + cigs
 algorithm:    sampling
 sample:       8000 (posterior sample size)
 priors:       see help('prior_summary')
 observations: 3140
 predictors:   8
 
plot(post_m1, "areas", prob = 0.95, prob_outer = 1, pars=c("age", "height", "weight", "sdp", "dbp", "chol", "cigs"))
\end{lstlisting}
Figure \ref{fig:example3} shows the marginal posterior distributions of the regression coefficients $\beta_j$ for the Bayesian logistic regression model in Example 3.
To compute the FBST on the regression coefficients, we need to extract the posterior MCMC sample first, as shown in listing \ref{r:listing14}. For illustration purposes, we conduct the FBST on the regression coefficient belonging to the covariate weight. The FBST is computed using a normal prior $\mathcal{N}(0,2.5)$ as reference function, which was also used to fit the model. This way, the surprise function quantifies which parameter values $\beta_j$ have been corroborated more by observing the data than the null value $\beta_j=0$.
\begin{lstlisting}[caption={Example 3 - Extracting the posterior MCMC draws, performing the FBST and visualising the result}\label{r:listing14},language=R]
    posteriorDrawsMatrix = as.matrix(post_m1)
    weightDraws = posteriorDrawsMatrix[,4]
    
    resWeight = fbst(posteriorDensityDraws = weightDraws, nullHypothesisValue = 0, dimensionTheta = 8, dimensionNullset = 7, FUN=dnorm, par = list(mean = 0, sd = 2.5))
    # Bayesian evidence against null hypothesis
    resWeight@eValue
    [1] 0.9758885
    # Standardized e-value
    resWeight@sev_H_0
    [1] 0.00002672151
        
    plot(resWeight)
\end{lstlisting}
The results are also shown in figure \ref{fig:example3fbst}, which is produced via the \texttt{plot()} function call in listing \ref{r:listing14}.
Based on the standardized $e$-value $\text{sev}(H_0)\approx 0.0000267$ and the Bayesian evidence against $H_0$, the $e$-value $\overline{\text{ev}}(H_0)\approx 0.9759$ one would reject the null hypothesis $H_0:\beta_j=0$.

\section*{Discussion}
This paper introduced the R package \texttt{fbst} for computing the Full Bayesian Significance Test and the $e$-value for testing a sharp hypothesis against the alternative. The conceptual approach and the statistical theory of the FBST were detailed, and three examples of statistical models frequently used in psychology and the cognitive sciences highlighted how the FBST can be computed in practice via the \texttt{fbst} R package. It was shown that both one-sided and two-sided hypotheses can be tested with the \texttt{fbst} package. The package's core function \texttt{fbst()} requires only a posterior MCMC sample so it should be applicable to a wide range of statistical models used in the cognitive sciences and psychology. The examples demonstrated that it is simple to combine the FBST via the \texttt{fbst} package with widely used libraries like \texttt{rstanarm} \citep{Goodrich2020} or the \texttt{BayesFactor} package \citep{BayesFactorPackage}. The provided summary and plot functions in the package allow intuitive use and produce appealing visualisation of the FBST results which simplifies sharing and communication of the results with colleagues. We omitted simulation studies in this paper, because these were recently conducted by \cite{Kelter2020BayesianPosteriorIndices} to which the interested reader is referred. For more details on the theoretical properties of the FBST, we also refer the reader to \cite{Pereira2020}.

To conclude, we direct attention to some limitations and possible extensions of the FBST and the \texttt{fbst} package presented in this paper. First, the \texttt{fbst} package is widely applicable but this strength can also be interpreted as a limitation. The \texttt{fbst} package requires a posterior distribution which has been derived analytically or numerically to conduct the FBST and compute the $e$-value, so it is not a standalone solution.

Second, the core functionality in the current form is restricted to computing, summarising and visualising the FBST. Future extensions could include more detailed analysis results like robustness checks depending on the reference function used, see \cite{VanDoorn2019}. Also, in its current form the package uses only posterior MCMC draws, and future versions could provide the option to provide the posterior as a closed-form function. Another option to extend the functionality would be to make various algorithms available to estimate the posterior density based on the posterior draws: By now, only Gaussian kernel density estimation is used. In small sample situations the asymptotics of Bayesian posterior distributions guaranteed by the Bernstein-von-Mises theorem can be questionable and other approaches like spline-based interpolation or non-Gaussian kernels may be more useful.

Third, while the standardized $e$-values may be used as a replacement of frequentist $p$-values, they are also based on asymptotic arguments and future research is needed to study the behaviour of the standardized $e$-values $\text{sev}(H_0)$ for small samples. This is why we recommend a continuous interpretation of the Bayesian $e$-value $\overline{\text{ev}}(H_0)$ over a threshold-oriented interpretation via standardized $e$-values $\text{sev}(H_0)$.

In closing, it must be emphasized that we do not argue against the appropriate use of $p$-values, Bayes factors or any other suitable method of hypothesis testing. However, the ongoing debate about the concept of statistical significance shows that it is useful to explore existing alternatives for statistical hypothesis testing and investigate the relationships between these approaches both from a theoretical and practical perspective \citep{Berger1987,Makowski2019,Liao2020}. The \texttt{fbst} R package introduced in this paper could contribute in particular to the former, as simulation studies can easily be carried out by employing the package, see for example \cite{Kelter2020BayesianPosteriorIndices}.

There is much value in testing a sharp null hypothesis against its alternative in the cognitive sciences and psychology \citep{BergerBrownWolpert1994,Berger1997,Rouder2009}. While there are also other useful approaches such as equivalence testing -- see \cite{Lakens2017,Lakens2018,Kruschke2018,Kruschke2018a} -- the FBST has shown to be an attractive alternative to NHST and $p$-values with desirable theoretical and practical properties \citep{Kelter2020BayesianPosteriorIndices,Pereira2020,Esteves2019}. It is hoped that this package will be useful to researchers from the cognitive sciences and psychologists who are interested in a fully Bayesian alternative to null hypothesis significance testing which requires little methodological changes, but offers all the benefits of a fully Bayesian data analysis.

\section*{Conflict of interest}
The authors declare that they have no conflict of interest.

\section*{Open Practices}
The data and materials for all analyses are available at \url{https://osf.io/u6xnc/}.

\bibliographystyle{apa}
\bibliography{library.bib}

\end{document}